\documentclass[12pt]{article}
\usepackage{fullpage}
\usepackage{graphicx}
\usepackage{float}
\usepackage{amsmath}
\usepackage{amssymb}
\usepackage{amsthm}
\usepackage{url}
\usepackage[hidelinks]{hyperref}
\usepackage{color}
\usepackage{threeparttable}
\usepackage[hmargin=1.0in,vmargin=1.0in]{geometry}
\usepackage{authblk}

\usepackage[markup=underlined]{changes}
\usepackage{todonotes}
\makeatletter
\setremarkmarkup{\todo[color=Changes@Color#1!20,size=\scriptsize]{#1: #2}}
\makeatother

\definechangesauthor[color=blue]{DJ}
\definechangesauthor[color=red]{MTC}
\definechangesauthor[color=green]{HI}

\usepackage{fancyhdr}
\usepackage{latexsym}
\usepackage{verbatim}
\usepackage{multirow}
\usepackage{caption} 
\usepackage{natbib}
\usepackage{listings}
\usepackage{lineno}
\usepackage{setspace}

\begin{document}
\title{{\bf Statistical methods for the quantitative genetic analysis of high-throughput phenotyping data}}

\author[1]{Gota Morota}
\author[2]{Diego Jarquin}
\author[1]{Malachy T. Campbell}
\author[3]{Hiroyoshi Iwata}
\affil[1]{Department of Animal and Poultry Sciences, Virginia Polytechnic Institute and State University, Blacksburg, VA, USA 24061}
\affil[2]{Department of Agronomy and Horticulture, University of Nebraska-Lincoln, Lincoln, NE, USA 68583}
\affil[3]{Department of Agricultural and Environmental Biology, Graduate School of Agricultural and Life Sciences, The University of Tokyo, Tokyo, Japan 113-8657}

\date{}

\maketitle

\newpage 

\noindent 
Corresponding author: \\
Gota Morota \\
Department of Animal and Poultry Sciences \\
Virginia Polytechnic Institute and State University \\
Blacksburg, Virginia 24061, USA \\
E-mail: morota@vt.edu \\

\newpage
\doublespacing
\section*{Abstract} 
The advent of plant phenomics, coupled with the wealth of genotypic data generated by next-generation sequencing technologies, provides exciting new resources for investigations into and improvement of complex traits. However, these new technologies also bring new challenges in quantitative genetics, namely, a need for the development of robust frameworks that can accommodate these high-dimensional data. In this chapter, we describe methods for the statistical analysis of high-throughput phenotyping (HTP) data with the goal of enhancing the prediction accuracy of genomic selection (GS). Following the Introduction in Section 1, Section 2 discusses field-based HTP, including the use of unmanned aerial vehicles and light detection and ranging, as well as how we can achieve increased genetic gain by utilizing image data derived from HTP. Section 3 considers extending commonly used GS models to integrate HTP data as covariates associated with the principal trait response, such as yield. Particular focus is placed on single-trait, multi-trait, and genotype by environment interaction models. One unique aspect of HTP data is that phenomics platforms often produce large-scale data with high spatial and temporal resolution for capturing dynamic growth, development, and stress responses. Section 4 discusses the utility of a random regression model for performing longitudinal GS. The chapter concludes with a discussion of some standing issues.

\newpage
\section{Introduction}
The predicted rise in global temperatures, increased variability of precipitation events, and increased competition for freshwater resources and arable land threaten to place unique constraints on global agriculture. Plant breeders in the 21\textsuperscript{st} century will need to develop cultivars that are both high-yielding and resilient to climate change. The evaluation and development of breeding material requires a multifaceted approach, necessitating the consideration of multiple complex, and often interdependent traits. Successful germplasm development is not only dependent on the increase in the performance of breeding material that is achieved each cycle but also the amount of time before a cultivar is released to the end-users. Moreover, the development of elite cultivars tolerant to abiotic stresses requires careful consideration of a suite of morphological and physiological traits that facilitate adaptation (e.g., plasticity and stability) to a range of environmental conditions. Thus, genetic improvement in this respect is a highly demanding process that requires extensive phenotypic evaluation in multiple environments. Advancements in sequencing have led to new genomic tools and have opened new avenues of research that aid breeders in their selection procedure. For instance, next-generation sequencing (NGS) techniques such as genotyping-by-sequencing \citep{elshire2011robust} have significantly increased the number of markers discovered and the number of individuals that can be sequenced, providing a cost-efficient tool for breeders to obtain the genotypic profiles of individuals.

In parallel with NGS advancements, new statistical methods have been developed to enable utilization of the vast amount of available genomic information for selection purposes. This is known as genomic selection (GS), and its fundamental concept was first introduced by \citet{meuwissen2001prediction}. GS predicts the performance of unobserved individuals based on the linkage disequilibrium (LD) between markers and causal loci and the genomic relatedness between observed and unobserved individuals. It has been shown that GS can increase genetic gain by reducing the number of cycles and the number of progeny that need to be phenotypically tested, thus reducing the cost of a breeding pipeline. Since \citet{meuwissen2001prediction}, there have been improvements in the prediction accuracy of GS through the incorporation of pedigree information \citep{de2009predicting,crossa2010prediction,aguilar2010hot,christensen2010genomic}, environmental covariates, and genotype by environment interactions \citep{burgueno2012genomic,heslot2014integrating,jarquin2014reaction,perez2017single}. 

One of the main advantages of GS over phenotypic selection is that phenotypic information is not required for the validation set. However, the acquisition of accurate phenotypic information is still a crucial component for the training or calibration set in the model building process. In other words, the phenotypic information of selection candidates is not used directly for selection, but the predictive ability of the models is negatively affected by the absence of accurate phenotypes. Obtaining precise phenotypic values is not trivial, but it is a critical part of genome-based breeding  \citep{white2012field,cobb2013next,crain2018combining}.

In recent years, high-throughput phenotyping (HTP) has become an emerging technology that can assist breeders in improving selection procedures and developing commercial cultivars more rapidly and efficiently \citep{furbank2011phenomics}. In particular, image-based plant phenotyping enables frequent, non-destructive evaluation of multiple traits for a large number of plants with high precision. Image-based phenotyping offers several advantages, including being generally non-destructive, requiring low or no physical human labor input, being cost effective, and the ability to measure multiple traits at the same time in different locations and at different developmental growth stages \citep{white2012field,reynolds2018cost}. 

There are a wide range of HTP platforms that have been developed for the purpose of providing dense phenotypic information \citep{fiorani2013future}. Remote sensing and robotic systems developed in greenhouses and growth chambers have a high initial cost but can be fully automated. Alternatively, HTP data can also be collected in the field, as described later. In general, these field-based systems are associated with a high initial cost and also require a well-trained operator for collecting high-quality data \citep{reynolds2018cost}. \citet{reynolds2018cost} characterized and compared the available platforms in terms of associated costs and purposes. Despite such costs, a growing number of breeding programs are utilizing HTP platforms to better understand the genetics of quantitative traits and leverage these high-dimensional data to enhance selection. Specifically, HTP can be used to both generate dependent phenotypic variables for the training set in prediction models and provide additional information on genetic predictor variables in GS models, thereby improving prediction accuracies for conventional breeding targets, such as yield. Thus, this type of data can serve two main purposes: 1) as a primary trait response (e.g., plant height, canopy coverage, and number of leaves), and 2) as a covariate associated with the target trait response (e.g., yield). We discuss these points further in the following sections.

\section{Field-based high-throughput phenotyping using UAV}
In this section, we show that HTP accelerates plant breeding by improving the response to selection \citep{araus2018translating}. HTP methods allow us to measure plants efficiently and accurately via automatic or semi-automatic analysis of data collected by cameras and sensors \citep{fahlgren2015lights}. Methods for measuring plants cultivated in a field are collectively known as field-based HTP. Field-based HTP enables the measurement of a large number of plots in an experimental field using cameras and sensors mounted on different platforms, such as unmanned aerial vehicles (UAV) \citep{liebisch2015remote}, carts \citep{white2013flexible}, tractors \citep{andrade2014development}, and gantry cranes. Field-based HTP not only improves the efficiency and accuracy of phenotyping of plants in a field but also makes it possible to evaluate traits that are difficult to measure with conventional phenotyping methods. In particular, the UAV is one of the most cost-effective and easy-to-use platforms \citep{sankaran2015low,yang2017unmanned}. Although the type of camera or sensor that can be mounted on a UAV is restricted by the payload capacity of the UAV, light-weight and small cameras and sensors have been developed, and their precision and cost-efficiency have rapidly improved in recent years. The UAV is commonly equipped with digital cameras, multispectral cameras, and thermal infrared imagers in field-based HTP \citep[e.g.,][]{tattaris2016direct,kefauver2017comparative}. In contrast, hyperspectral cameras and LiDAR (light detection and ranging) are currently not commonly mounted on a UAV but are mainly ground-based HTP platforms \citep{madec2017high,yu2018hyperspectral,yuan2018wheat,sun2018field,wang2018field} because of their weight, size, and cost. The commercialization of the UAV and related equipment is progressing in various fields, and various measurement devices will be available for HTP in the near future.

Plant characteristics that can be measured using UAV are roughly divided into three types of traits: 1) geometric traits, 2) spectral traits, and 3) physiological traits. For geometric traits, plant height, canopy cover, and canopy volume are measured mainly with RGB cameras or multispectral cameras \citep{shi2016unmanned,madec2017high,watanabe2017high,han2018clustering,li2018elucidating,pugh2018temporal,spindel2018association,sun2018field,wang2018field,yuan2018wheat}. To measure these traits, a method called Structure from Motion (SfM) is used to estimate the three-dimensional (3D) structures of plants or plant canopies from a sequence of images acquired by a UAV. The structure is obtained using a set of data points, called a point cloud, in a 3D space. The 3D coordinate information of a point cloud is converted into a digital surface model (DSM) and an orthomosaic image. DSM is used for measuring plant 3D structural traits, such as plant height and canopy volume, while orthomosaic images are used for traits evaluated from above the ground, such as canopy cover. Lodging of plants can also be measured by DSM analysis \citep{chapman2014pheno}. The numbers and locations of flowers, blooms, and heads are also measured as geometric traits \citep{guo2018aerial,xu2018aerial}. In these studies, image-based machine learning has been used for the detection of target objects (i.e., flowers, blooms, or heads) from images acquired by UAV. \citet{guo2018aerial} employed a two-step machine learning method for the detection of sorghum heads and attained high accuracy on various genotypes with different head morphologies and at different growth stages. \citet{xu2018aerial} used a convolutional neural network to detect cotton blooms and estimated the 3D coordinates of the blooms using a dense point cloud constructed by SfM. These two studies demonstrated the potential of the combinatory use of image-based machine learning and HTP. Moreover, these studies suggest that simple but labor-intensive measurements, such as the monitoring of flowering and heading, can be performed on a much larger scale with HTP and image-based machine learning than with conventional methods. 

For spectral traits, vegetation indices (VI) calculated from multispectral images acquired by UAV are used for evaluating vegetation properties, such as plant structure, biochemistry, and plant physiological and stress status \citep{haghighattalab2016application,zaman2015unmanned,condorelli2018comparative,shi2016unmanned,li2018elucidating,potgieter2017multi,kyratzis2017assessment,hassan2018time,vergara2016novel,jay2018exploiting,verger2014green,han2018clustering}. A large number of VIs have been proposed and have been used in ground-based platforms, aircraft, and satellite remote sensing. The fine spatial resolution of a UAV enables the removal of soil and shadow pixels from images and can improve the estimation of vegetative properties. 
\citet{jay2018exploiting} used 6-band multispectral cameras to evaluate the structural and biochemical plant traits of green fraction, green area index, leaf chlorophyll content, and canopy chlorophyll and nitrogen contents, showing that the fine spatial resolution of the UAV always improved the estimation accuracy of these traits. Although multispectral images allow us to estimate various VIs better than RGB images, multispectral cameras are usually more expensive and have lower resolution than RGB cameras. To resolve this issue, \citet{khan2018estimation} proposed a method for model-based estimation of VIs using RGB images. In this method, mean VI values were computed from the near infrared and red channels of corresponding plots, and then a deep neural network was trained with the RGB images as the input source and the VI values as the target output. A similar approach can be applied to the estimation of hyperspectral VIs from multispectral or RGB images and will be useful because hyperspectral cameras are usually much more expensive than multispectral cameras. 

As for physiological traits, traits such as leaf chlorophyll content, protein content, biomass, crop vigor, nutrition status, and water status are measured by various methods including 3D construction and spectral VIs. A method that is specific to physiological traits is thermal infrared imaging, which enables the measurement of canopy temperature and can be used as a tool to indirectly evaluate the transpiration rate of a plant. \citet{tattaris2016direct} used a thermal infrared camera and a multispectral camera for HTP, using UAV to measure canopy temperature and the VI of wheat, and found that data acquired by UAV generally exhibited stronger correlations with yield and biomass than data obtained from ground-based phenotyping. \citet{ludovisi2017uav} applied thermal infrared imaging to measure the canopy temperature of black poplar using UAV and found that the canopy temperature showed a good correlation with ground-truth stomatal conductance. Although the canopy temperature is an important indicator of stress status, it is extremely sensitive to small environmental changes, making it difficult to assess through slow, ground-based methods \citep{chapman2014pheno}. HTP using UAV provides a good solution for this problem.

\subsection{Application of HTP in breeding populations}
When selecting breeding populations using HTP, two relatively simple methods are considered: 1) indirect selection and 2) index selection. Another method, selection based on prediction with HTP and genomic information, will be described later. When genetic correlations exist, selection for one trait will cause corresponding changes in other traits that are correlated \citep{acquaah2009principles}. This change in response due to genetic correlation is called a correlated response and may be caused by pleiotropy or LD. 

In the indirect selection method, a target trait, X, is selected indirectly by selection for trait Y, which has a genetic correlation with trait X and can be measured by HTP. It is possible to improve the selection efficiency of the target trait, the measurement of which would be costly, time consuming, or labor intensive, with traits readily measured by HTP. For example, \citet{madec2017high} measured wheat plant height with HTP using LiDAR and UAV and found that it was highly correlated with the plant height measured at the ground level. They also demonstrated that heading date could be estimated based on a growth curve of plant height measured by LiDAR. \citet{kyratzis2017assessment} evaluated the potential use of VIs acquired by UAV for durum wheat phenotyping and found that one index was significantly correlated with grain yield. 

In the index selection method, a target trait, X, is selected based on an index calculated from phenotypes of a set of $m$ traits, Ys, related to the target trait. The simplest index is a linear combination expressed as
$$
I = \sum^m_{j=1}b_j y_j
$$
where $b_j$ and $y_j$ are the weight and phenotypic value of trait $\text{Y}_j$, respectively. If we consider $b_j$ and $y_j$ as the effect and state of marker $j$ in a set of genome-wide markers, index selection becomes GS. The weight, which represents the relative importance of each trait, can be determined by multivariate regression. For example, \citet{kefauver2017comparative} built a model regressing the grain yield on VIs acquired by HTP using UAV with stepwise regression and found that the regression model explained 77.8\% of the grain yield variation. 
\citet{yu2018hyperspectral} performed hyperspectral imaging of a wheat canopy and used the resulting data to detect \textit{Spectoria tritici} blotch disease and to quantify the severity of infection. They used partial least squares regression to build a prediction model for severity and found that the accuracy of prediction (correlation between observed and predicted values) was 0.38-0.60 for three disease metrics. Non-linear relationships between trait X and a set of traits Ys can also be modeled in a selection index. Various types of models, including known and ad hoc machine learning models, can be used for building an index. \citet{thorp2018high} proposed a method for deriving daily evapotranspiration based on a daily soil water balance model named FAO-56 \citep{allen1998crop}, which was derived from an index acquired by HTP using UAV, to evaluate and improve the crop water use efficiency of cotton varieties. Collectively, indirect or index selection based on traits measured by HTP has strong potential to streamline the selection of important agronomic traits such as plant height, heading date, grain yield, and disease resistance.

\subsection{Genetic gain in HTP-based selection}
HTP-based selection and GS can accelerate plant breeding by improving the efficiency of selection. Response to selection is an index for evaluating the efficiency of selection \citep{falconer1996introduction}. The response to selection $R$ is defined as the difference between the mean phenotypic values ($\bar{y}_{o}$) of progeny generated from the selected parents and the mean phenotypic values ($\bar{y}_p$) of the parental population before selection.
$$
R = \bar{y}_{o} - \bar{y}_{p}.
$$
If we denote the heritability of a trait targeted in the selection as $h^2$ and define the selection differential as the product of the phenotypic standard deviation $\sigma_p$ and selection intensity $i$ in the parent population,
$$
R = i h^2 \sigma_p.
$$
This is an important formula in breeding known as the ``breeder's equation". If a breeder knows the heritability of the target trait $h^2$ and the standard deviation of the phenotype $\sigma_p$ in the parent population, it is possible to calculate the expected response to selection $R$ under intensity $i$. Using the definition of heritability, $h^2=\sigma_g^2 / \sigma_p^2$, we can rewrite the formula as
$$
R=i h \sigma_g,
$$
where $\sigma_g$ is the square root of the genetic variance in the parent population.

Now we consider the case in which we select trait X indirectly by selecting for trait Y, measured with HTP. In this case, the response to selection of the indirect selection of trait X with trait Y is
$$
R_{XY} = i_Y h_Y r_{XY} \sigma_{gX},
$$
where $i_Y$ is the selection intensity of trait Y, $h_Y$ is the square root of the heritability of trait Y, $r_{XY}$ is the genetic correlation between trait X and trait Y, and $\sigma_{gX}$ is the square root of the genetic variance of trait X in the parent population.
To improve the efficiency of selection with HTP, this value should be larger than the response to selection of direct selection of trait X, i.e., $R_X = i_X h_X \sigma_{gX}$. That is, the condition for improving the selection efficiency with HTP is
$$
i_Y h_Y r_{XY} > i_X h_X.
$$

When the selection intensities of the two traits are the same ($i_Y = i_X$), the following two conditions should be satisfied: 1) trait Y measured by HTP has a higher heritability than trait X, and 2) the genetic correlation between trait X and trait Y is high. With HTP, however, it is often possible to evaluate a large number of genotypes (strains or individuals) as compared with direct selection of trait X using a conventional phenotyping method. Therefore, the selection intensity of trait Y can be increased compared to the selection intensity of trait X. If $i_Y > i_X$, even when the heritability of trait Y is not larger than that of trait X, it may be possible to perform indirect selection on trait X with higher efficiency than that of direct selection. 

Index selection with HTP and GS both involve indirect selection of trait X based on the index I, which is calculated based on traits measured by HTP or genome-wide marker genotypes. The response to selection is represented as
$$
R_{XI} = i_I r_{XI} \sigma_{gX},
$$
where $i_I$ is the selection intensity of the index I and $r_{XI}$ is the accuracy of selection of trait X based on index I. The condition that the response to index selection is greater than the response to direct selection of trait X is
$$
i_I r_{XI} > i_X h_X.
$$

When the selection intensities of index I and trait X are the same ($i_I = i_X$) and the accuracy $r_{XI}$ of selection of trait X based on index I exceeds the square root of the heritability of trait X, $h_X$, the efficiency of selection by index selection exceeds the efficiency of direct selection of trait X. As in the case of indirect selection using trait Y, if $i_I > i_X$, even when the accuracy $r_{XI}$ of selection of trait X based on index I does not exceed the square root of the heritability of trait X, index selection has a higher efficiency than direct selection. 

When we consider the efficiency of a breeding program, it is important to evaluate the genetic gain per unit time. Dividing the reaction to selection $R$ by the time $\delta_X$ required for one cycle of selection, we obtain 
$$
\Delta G_X =\frac{i h_X \sigma_{gX}}{\delta_X},
$$
where $\Delta G$ is the genetic gain per time. The genetic gain of indirect selection of trait X with trait Y is
$$
\Delta G_{XY} =\frac{i h_Y r_{XY} \sigma_{gX}}{\delta_Y},
$$
and the genetic gain of index selection of trait X with index I is
$$
\Delta G_{XI} =\frac{i_{I}  r_{XI} \sigma_{gX}}{\delta_I}.
$$
Here, $\delta_Y$ and $\delta_I$ are the times required for one cycle of indirect and index selection, respectively. The time required for one cycle of selection can be shorter for trait Y and index I than for trait X. For example, the yield and quality of a grain crop are usually evaluated with multiple plants on a plot-by-plot basis. However, in indirect and index selection, it may be possible to perform selection on a single plant basis in earlier generations, such as second-generation hybrids (F$_2$). 
In such a case, $\delta_Y$ (or $\delta_I$) $< \delta_X$, and even when the response to selection $R_{XY}$ or $R_{XI}$ is smaller than the response to selection $R_{X}$, the genetic gain per unit time becomes greater under indirect and index selection than under direct selection. 

As described above, the efficiency of selection can be improved by taking advantage of HTP, especially in terms of improvements in selection intensity and the time required for one cycle of selection. Field-based HTP is useful for increasing selection intensity because of its scalability, while HTP in the greenhouse is good for reducing the time required for one cycle of selection because it is often performed on a single-plant basis and year-round. In the application of HTP in plant breeding, the factors described earlier should be taken into account to optimize selection methods for target traits.

\subsection{Use of HTP for GWAS and GS}
Although HTP alone is expected to improve the response to selection, response to selection can be further improved by using HTP in combination with genome-wide association study (GWAS) and GS. HTP with UAV is particularly suited for this purpose, as it can measure a large number of small- to medium-sized plots in which plants are cultivated. HTP with UAV has been applied to the evaluation of a large number of genotypes (germplasm accessions, varieties, and breeding lines) in many species, including wheat \citep{sankaran2015field,haghighattalab2016application,rutkoski2016canopy,madec2017high,condorelli2018comparative}, maize \citep{shi2016unmanned,han2018clustering}, sorghum \citep{watanabe2017high,guo2018aerial,spindel2018association}, and black poplar \citep{ludovisi2017uav}. \citet{condorelli2018comparative} performed GWAS with 248 elite durum wheat lines to compare the results obtained with two UAVs and a ground-based method to measure a VI (Normalized Difference Vegetation Index, NDVI). More associations were detected by HTP using UAV than with the ground-based method, suggesting an improved ability of HTP using UAV over the ground-based method. \citet{spindel2018association} undertook GWAS with 648 diverse sorghum lines for 460 combinations of traits, treatments, time-points, and locations. Four traits related to biomass, plant height, and leaf area were measured by HTP using UAV. In total, 213 high-quality, replicated, and conserved associations were detected in genomic intervals, including many strong candidate genes. \citet{watanabe2017high} measured the height of 115 sorghum germplasm accessions with HTP using UAV and evaluated the potential of HTP to provide phenotypic training data in a GS model. Although phenotypic correlation was not high, GS of plant height as measured by HTP using UAV was highly correlated with those measured manually. These results suggest the considerable potential of HTP using UAV for genomic-assisted breeding through GWAS and GS. 

To successfully combine HTP with GWAS or GS, a novel viewpoint different from the analysis of conventional phenotypic data is necessary. Since HTP enables non-destructive and frequent measurements for large-scale field tests, a target trait can be measured as high-density time series data and as high-density data with coordinate information. Thus, spatial-temporal continuity and change can be taken into account in GWAS and GS models. For instance, \citet{elias2018improving} fitted a model with a spatial kernel as well as a kernel-based genomic relationship matrix to cassava agronomic trait data to account for the spatial heterogeneity in the field and showed that the prediction accuracy increased after accounting for the spatial variation. Moreover, multiple sensors are commonly employed in HTP, each of which can acquire high-dimensional data (e.g., hyperspectral images). Thus, for GWAS and GS using phenotypic data collected by HTP, it is necessary to consider the high-dimensionality of the data and the large number of data points. \citet{spindel2018association} conducted a GWAS on a number of features collected with HTP using UAV and constructed a method and pipeline to fuse and organize numerous GWAS results. Phenotypic data measured by HTP can also be used in the prediction of genotypic values of a target trait by leveraging genetic correlations between the target trait and traits measured by HTP. \citet{rutkoski2016canopy} proposed a method for predicting a target trait with correlated HTP traits, as described in the next section.

\section{Integration of HTP data into GS}
\subsection{Single-trait analysis}
Recently, there have been several studies that have integrated genomic data and HTP data for prediction purposes in several crops using different modeling techniques \citep{rutkoski2016canopy,xavier2017genetic,sun2017multitrait,cabrera2012high,krause2018use,crain2018combining,juliana2019integrating,jarquin2018increasing}. The integration of genomic and HTP data provides opportunities to improve existing GS models, thus enabling breeders to select their material more accurately and increase genetic gain. We summarize some key methods developed for integrating high-throughput genomic and HTP information for the purpose of increasing the accuracy of prediction by extending the standard GS models. 

We can include secondary image traits in a quantitative genetics model using two model parameterizations. The first model explains the $i$th phenotypic observation as the sum of an intercept $\mu$ common to all observations, a linear combination of $p$ markers $x_{ij}$ and their corresponding marker effects $b_j$, a linear combination of $Q$ secondary traits $s_{iq}$ and their corresponding effects $a_q$, and residual $\epsilon_i$ as follows: 
$$
y_i = \mu + \sum^p_{j=1}x_{ij}b_j +  \sum^Q_{q=1}s_{iq}a_q + \epsilon_i .
$$
The second model parameterization is based on covariance structures and can be obtained from the previous model by assuming that the effects of marker $b_j$ and secondary traits $a_q$ are independent and identically distributed draws from normal densities of the form $b_j \sim N(0, \sigma^2_b)$ and $a_q \sim N(0, \sigma^2_a)$. Then, $g_i = \sum^p_{j=1}x_{ij}b_j$ and $w_i = \sum^Q_{q=1}s_{iq}a_q$ are genetic and environmental values of the $i$th genotype using information from genomics and secondary traits. From properties of the multivariate normal density, the vectors of marker and secondary trait effects are also normally distributed, such as $\mathbf{g} =\{g_i\} \sim N(\mathbf{0}, \mathbf{G}\sigma^2_{g})$ and $\mathbf{w} = \{w_i\} \sim N(\mathbf{0}, \mathbf{C}\sigma^2_{A})$, where $\mathbf{G} = \mathbf{XX}'/p$ is a covariance matrix whose entries describe genomic similarities between pairs of genotypes;  $\mathbf{X}$ is the matrix of molecular markers of order $n \times p$; $\sigma^2_{g} = p  \times \sigma^2_{b}$; $\mathbf{C} = \mathbf{SS}'/Q$ is a covariance matrix whose entries describe phenotypic similarities based on image secondary traits data for each pair of observations;  $\mathbf{S}$ is the matrix of secondary traits of order $n \times Q$; and $\sigma^2_{A} = Q  \times \sigma^2_{a}$. This parameterization assumes that all of the secondary traits equally contribute to explain the phenotypic variations of the traits of interest. One of the advantages of using this second parameterization is that it is possible to evaluate the contribution of the genomic and HTP components for explaining phenotypic variability by comparing the estimated variance components associated with each of these terms.

The majority of models developed focus on predicting a single trait, namely, grain yield. HTP can measure traits that are shown to be highly correlated with grain yield, such as the spectral reflectance of the canopy and canopy temperature \citep{amani1996canopy}. A VI is used to summarize the spectral reflectance of the canopy scores \citep{krause2018use}. However, because the VI is calculated using only a subset of the available wavelengths, it does not take advantage of all of the HTP data. There are several approaches for incorporating all of the HTP wavelengths and the plot-level VI measurements into GS models. \citet{rutkoski2016canopy} showed that the integration of VI and canopy temperature into a genomic best linear unbiased prediction (GBLUP) model could increase the prediction accuracy by 70\% compared to that of a univariate baseline model in wheat data. \citet{aguate2017use} showed that using bands as predictors increased prediction accuracy over that of VI. They used ordinary least squares, partial least squares, and a Bayesian shrinkage model to incorporate wavelengths into a GS model in maize.
A similar observation was made by \citet{montesinos2017predicting}, who compared prediction model performance when all of the wavelengths were incorporated with that of a subset of the wavelengths in wheat. They  concluded that using all of the wavelengths resulted in higher  prediction accuracy.

\subsection{Multi-trait analysis}
\citet{sun2017multitrait} predicted grain yield in a two-step procedure in wheat data.  First, they collected data on canopy temperature and VI as secondary traits (which are correlated with grain yield) and modeled the secondary traits using the genetic marker and environmental effects. They applied a mixed model for predicting grain yield without considering the secondary traits as covariates. However, they used the secondary traits to develop a multivariate model to predict grain yield, which is the primary trait. The secondary traits were measured in a longitudinal fashion, i.e., at several time points throughout the growing season. They implemented and compared the repeatability, multi-trait, and random regression (RR) models that can be used for modeling longitudinal data. In the second step of the GS, the results from the repeatability, multi-trait, and RR models were used as BLUP, and a univariate prediction model was compared to bivariate and multivariate models. Only grain yield was included, and the secondary traits were excluded in the univariate model. In one of the multivariate prediction models, the secondary traits were included both in the training and testing sets, and in the other multivariate prediction model the secondary traits were included only in the training set. The bivariate prediction model included grain yield and one of the secondary traits. Their results showed that the multivariate prediction model that incorporated the secondary traits in both the testing and training sets had an advantage over the other models in terms of prediction accuracy. However, it was not clear which of the first models (repeatability, multi-trait, or RR) performed the best because the results depended on the environmental conditions. Nonetheless, the results clearly demonstrated the advantage of using HTP data in GS applications. 

\citet{crain2018combining} compared four models using wheat data: 1) a regular GS model, 2) a univariate model in which grain yield was the response and HTP data were predictors, 3) a model which was the combination of models 1 and 2, and 4) a multi-trait model that included grain yield, canopy temperature, and VI measurements. The results showed that the addition of HTP data increased the prediction accuracy. They found that the multi-trait model exhibited a 7\% gain in terms of prediction accuracy, indicating that collecting multiple HTP measurements has the potential to increase genetic gain through the improvement of prediction models. \citet{juliana2019integrating} applied multivariate prediction models to compare standard GS with a pedigree- and HTP-based prediction model. They discussed the situations in which each model can be useful and the importance of implementing the correct models in the correct stage of the breeding pipeline. The authors elaborated on the importance of the family structure and of the secondary HTP traits being highly correlated with the primary phenotypic trait, as these components are influencing factors in prediction performance. 

\subsection{Genotype by environment interactions}
Although all of the studies described above considered approaches to integrate HTP into GS, they did not apply interaction effect models.  However, there are multiple lines of evidence that GS models with interaction effects have the potential to outperform competing models with only additive effects \citep{roorkiwal2018genomic,sukumaran2017pedigree,jarquin2017increasing}. \citet{montesinos2017genomic} presented one of the first studies of HTP showing the impact of including the interaction between hyperspectral bands and environment (band $\times$ environment). These authors found that the model with the band $\times$ environment interaction outperformed all of the models without this interaction term. \citet{jarquin2018increasing} used prediction models that incorporated line, environment, marker genotype, canopy coverage image information, and their interactions in soybean. They evaluated six main effects models that included combinations of line, environment, marker genotype, and canopy coverage image information; seven models with two-way interactions among the components; and two models with a three-way interaction between environment, marker genotype, and the canopy coverage data. Under the GBLUP model, they modeled the interaction components as the Hadamard product \citep{davis1962norm} of the relationship matrices obtained from genetic marker and canopy coverage image information according to the reaction norm model \citep{jarquin2014reaction}. The model performance was evaluated using three cross-validation (CV) schemes: CV2, CV1, and CV0. CV2 assumed an incomplete field trial, in which some lines are observed in some environments but not in others. CV1 was the case in which one predicts the performance of a new line in environments in which some other lines were evaluated. The goal of CV0 was to predict the performance of already tested lines in untested environments. When grain yield was the target trait, the advantage of including the canopy coverage measurements and the interactions among marker, environment, and canopy coverage measurement was clearly shown. The highest predictive abilities for CV2 and CV1 were delivered by the models that included a three-way interaction among marker genotype, canopy coverage image data, and environment, while for CV0, the model with interactions between marker genotype and environment, and between canopy coverage image information and environment produced, the greatest accuracy. The study also evaluated the effectiveness of canopy coverage image data from early stages and compared it with the case in which the canopy coverage image data was collected throughout the growing season. The results indicated that the information collected in the early stages was sufficient for prediction and that the additional data collected in the later stages did not improve the prediction models significantly. The practical implication of this finding is important, as it shows that the same prediction accuracy can be achieved using fewer resources (time, measurements, and costs). 

\citet{krause2018use} used multi-kernel, multi-environment GBLUP models including genetic marker or pedigree, environmental, and hyperspectral band information for predicting grain yield in wheat. They found that when marker genotype or pedigree data are not available, the main effects model using the hyperspectral band data provided a similar accuracy of prediction compared to the main effects models including marker or pedigree information. Additionally, the model with interactions outperformed the main effects models. Their findings differed from those of \citet{jarquin2018increasing} with regard to the effectiveness of including partial HTP data. They concluded that the prediction accuracy increased when the HTP data from later stages were included. However, this difference is expected, as the crop development for wheat is significantly different from that for soybean. Finally, \citet{montesinos2017genomic} and \citet{montesinos2018bayesian} showed the advantages of performing functional analysis for reducing data dimensionality to extract a higher signal-to-noise ratio for each observed value. In addition, \citet{montesinos2017genomic} showed that when the HTP collected over multiple time points are combined using functional analysis, a small increase in prediction accuracy can be achieved relative to that of models that use data from a single time point.

\section{Utilizing image-derived longitudinal traits for genetic studies in plants}
The observable phenotype at a given time is the culmination of numerous biological processes that have occurred prior to observation. For example, consider a cereal such as wheat at maturity. The total above-ground biomass can be separated hierarchically into a number of distinct organs. The whole plant can be partitioned into main and auxiliary tillers, which can be further partitioned into leaf blades, leaf sheaths, and stems. This process can proceed further to lower organization levels, separating these organs into tissues and cellular components. At each level, the pattern timing of development is tightly controlled by complex genetic networks that, at the organ level, control the onset of primordial development and initiation of growth and, at the plant level, the transition from vegetative to reproductive development.

An additional layer of complexity is added to this when the effect of the environment on these processes is considered. The appearance of the plant at maturity is certainly a product of its genetic makeup; however the processes mentioned above are all tightly linked to the environment. The total size of the plant at maturity is a product of the resources (e.g., light, nutrients, and water) that were available throughout its life cycle. Plants need light and carbon dioxide to produce sugars through photosynthesis. Nutrients are combined with these sugars to generate nucleotides, proteins, and metabolites. Limitations on any of these inputs will slow or stunt growth. In addition to plant growth, the transition between developmental states is also linked directly to the environment. Several studies have shown that drought events can lead to earlier flowering and accelerated post-anthesis development (reviewed by \cite{shavrukov2017early}). Therefore, the phenotype is not a static entity. The observable phenotype is the result of dynamic genetic processes, the changing external environment, and the dynamic interplay between the two.

For most genetic applications, plants are often phenotyped at only one or a few time points. These phenotypes are an incredibly informative summation of the processes that have occurred over the life cycle of the plant, and they have been used quite successfully to select for a variety of complex traits. While for many applications these single time point phenotypes may be sufficient, they fail to capture the dynamic processes that have led to the observable phenotype. In most genetic studies, phenotypic evaluation is the largest, most time-consuming activity. Typical genetic studies consist of a mapping population with hundreds to thousands of individuals that are grown in replicates. Thus, for these studies, phenotyping at one time point is often a huge commitment, while evaluation at multiple time points is often unfeasible.  

In the last decade, the construction and accessibility of high-throughput phenotyping platforms have provided an attractive means for generating phenotypic data throughout the duration of a study in a non-destructive manner for a number of economically important crop species \citep{furbank2011phenomics,chapman2014pheno,zhang2016evaluation,watanabe2017high}. These platforms have been successfully deployed in controlled environments to quantify growth and physiological processes in response to drought and salinity \citep{berger2010high, campbell2015integrating}. Moreover, with the growing popularity of UAVs in the consumer market, a vast selection of hardware can be obtained at relatively low cost \citep{watanabe2017high}. These can be outfitted with various sensors or cameras and deployed routinely in the field to capture trait development over the growing seasons. In crop species, these temporal phenotypes have been used as covariates in genomic prediction frameworks to improve prediction accuracy for end-point phenotypes, such as yield \citep{sun2017multitrait, crain2018combining, jarquin2018increasing}. However, analysis of the longitudinal trait itself has been largely confined to genetic inference in crops species, while genomic prediction has been applied largely to perennial species \citep{apiolaza2000variance, apiolaza2001analysis, de2018genetic}. In the following section, we describe several approaches for genomic prediction of the longitudinal phenotype itself.

\subsection{Single time point genetic inference}
A seemingly straightforward approach for assessing dynamic genetic effects underlying longitudinal traits is performing linkage or association analysis at each time point independently \citep{wu1999time,yan1998molecular, wurschum2014mapping, moore2013high}. In one of the first applications of HTP for genetic studies in plants, \cite{moore2013high} used an image-based platform to quantify root gravitropic responses in \textsl{Arabidopsis} biparental mapping populations. The authors used a step-wise mapping approach at each time point to identify time-dependent quantitative trait loci (QTL) and used a post hoc approach to combine information on QTL detected across multiple time points. The post hoc approach effectively used two metrics to classify QTL into a persistent class, by averaging the LOD scores across time points, and transient QTL, by taking the maximum LOD across all time points. While this post hoc approach effectively combines statistics across time points and successfully classifies the temporal genetics of root gravitropism, the single time point mapping approach itself does not explicitly model the covariance across time points. Thus, the actual genetic inference step does not fully capture the phenotypic trajectories.

\subsection{Functional mapping}
Several other approaches have been proposed that directly consider the trait trajectories for genetic analyses. With these approaches, the trait values across all time points can be modeled using parametric or non-parametric mathematical functions. These models describe the phenotypic trajectories using a few parameters (for a review of parametric models in the context of plant growth, see \cite{paine2012fit}). Once an appropriate model has been chosen, genetic inference or prediction can proceed using a single-step or two-step approach.

\subsubsection{Single-step functional mapping} 
In the single-step functional mapping approach, model fitting and genetic analyses are performed within a single statistical framework. In the plant community, the single-step approach for functional genetic inference/mapping was first proposed by \cite{ma2002functional} to map QTL for stem diameter in \textsl{Populus}. Since then, the functional mapping approach has been applied to longitudinal traits in other species, such as humans and mice, and has been extended into the mixed model framework used for GWAS \citep{wu2003functional, wu2006functional, das2011dynamic, cui2006functional, he2010mapping}. The advantages of the single-step functional mapping approach is that it considers the full trait trajectories over time, yielding loci that influence the curve itself, and captures the covariance across time points, which should reduce residual variance and improve statistical power \citep{das2011dynamic}. Essentially, at each locus, the single-step functional mapping approach models the mean trajectories for each genotype and tests whether the time-dependent genetic effects are non-zero. 

There are two important considerations for the single-step functional mapping approach: 1) the choice of function to model the mean trajectories of each genotype, and 2) the appropriate residual covariance structure to account for the temporal nature of the data. The function to model the mean trajectories can be parametric or non-parametric and can be selected based on some prior knowledge of the phenotypic trajectories. For well-studied traits, such as growth, a number of parametric options exist, are biologically meaningful, and can be easily applied to the longitudinal dataset \citep{paine2012fit}. In cases in which no prior knowledge exists about the phenotypic trajectories, a nonparametric function, such as orthogonal Legendre polynomials or B-spline functions, can be utilized. The nonparametric functions are described in greater detail below. A number of covariance structures can be used to account for the temporal relationships among observations. The choice will be dependent on the balance between statistical efficiency and robustness. In the most robust case, the unstructured covariance matrix, the variance and covariance at each time point are unique and estimated from the data. While this places no constraints on the variance-covariances, the number of parameters that must be estimated can be prohibitively large for most studies. In many cases, simpler structures may be nearly as robust while estimating far fewer parameters.

\subsubsection{Two-step functional mapping}
In contrast to the single-step functional mapping approach, the two-step approach performs the model fitting and genetic analysis in two separate steps. First, the phenotypic trajectories are modeled for each individual, and the model parameters are used as derived traits for subsequent genetic analyses (e.g., GWAS, linkage analysis, or GS). This two-step approach has been successfully used to examine the genetic basis of rosette growth in \textsl{Arabidopsis} and for GWAS and GS of early vigor in rice \citep{bac2015genome,campbell2017comprehensive}. The advantages of this approach are that it is conceptually simple and easy to implement. Moreover, for most popular growth models, the parameters have biological meaning. For instance, growth can be modeled over the life cycle of the plant using a 3-parameter logistic function. Here, the inflection point can be calculated, which represents the transition from vegetative to reproductive growth. Thus, the researcher can select a specific attribute to study and select a specific model parameter that represents that attribute for analysis. Moreover, outside of genetic mapping, these parameters may provide biological insight into the a plant's phenotypic development. For instance, in \cite{campbell2017comprehensive}, the authors targeted a specific model parameter that described a plant's growth rate and showed that the plant hormone gibberellic acid may influence natural variation for the rate-controlling parameter. However, the major disadvantage of this method is that information is lost between the functional modeling and genetic analysis steps. Since environmental factors are not included in the functional modeling step, the residuals likely contain important information regarding non-genetic components of the phenotypic variance for the longitudinal phenotype.

\subsection{Insights from animal breeding for genomic prediction using longitudinal traits}
While the use of longitudinal phenotypes is relatively new in plant science, animal breeders have targeted longitudinal traits for decades \citep{schaeffer1994random}. In animal breeding, breeders are often interested in the development of a trait across an animal's life. For instance, in dairy cattle, test day milk yields are collected routinely. Moreover, other traits, such as feed intake, growth, and egg production \citep{schnyder2001genetic,luo2007estimation,baldi2010random,howard2015genome}, have also been examined in a longitudinal framework . With the extensive use of these traits in animal breeding, numerous frameworks have been well developed to accommodate the time axis and have been used extensively for inference on genetic and environmental variance components, as well as pedigree and GS. 

In the following subsection, we discuss several approaches that have been used for pedigree- or genomic-based prediction in animal breeding in a context that is applicable to plant breeding with HTP platforms. As mentioned above, a naive approach for GS using longitudinal data would be a univariate approach, in which a conventional mixed model is fitted at each time point. Here, we introduce the concept of longitudinal GS from a multivariate framework, as this is a relatively simple extension of the univariate approach, and extend these concepts to covariance functions and RR models that have been pioneered in animal breeding.

\subsection{Multivariate approaches for longitudinal genomic prediction}
To capture the covariance between time points, a logical progression from the univariate approach is to utilize a multivariate framework for longitudinal data. Thus, rather than considering the longitudinal trait as a consecutive series of measurements on the same trait, with the multivariate approach, we essentially ignore the order of the series and treat each time point as a separate trait. The multivariate framework allows each time point to have a unique variance and unique covariances between time points. The multivariate GS framework is well developed and has been widely utilized in both plant and animal systems. Moreover, the extension from the univariate approach is relatively straightforward.

Assume a simple case in which we are given three consecutive measurements for each individual. The model for each trait can be written as 
\begin{align}
\mathbf{y}_1 &= \mathbf{X}_1 \mathbf{b}_1 + \mathbf{Z}_1 \mathbf{u}_1 + \boldsymbol{\epsilon}_1 \\
\mathbf{y}_2 &= \mathbf{X}_2 \mathbf{b}_2 + \mathbf{Z}_2 \mathbf{u}_2 + \boldsymbol{\epsilon}_2 \\
\mathbf{y}_3 &= \mathbf{X}_3 \mathbf{b}_3 + \mathbf{Z}_3 \mathbf{u}_3 + \boldsymbol{\epsilon}_3 \\
\end{align}

where $\mathbf{y}_i$ is the vector of observations for trait $i$; $\mathbf{X}_i$ and $\mathbf{Z}_i$ are the incidence matrices for fixed effects and random effects, respectively, for trait $i$; $\mathbf{u}_i$ is the vector of random genetic effects for trait $i$; and $\boldsymbol{\epsilon}_i$ is the vector of residuals for trait $i$. Thus, the multivariate model is

\begin{align}
\begin{bmatrix}
\mathbf{y}_1 \\
\mathbf{y}_2 \\
\mathbf{y}_3 \\
\end{bmatrix}
&=
\begin{bmatrix}
\mathbf{X}_1 & 0 & 0 \\
0 & \mathbf{X}_2 & 0  \\
0 & 0 & \mathbf{X}_3 \\
\end{bmatrix}
\begin{bmatrix}
\mathbf{b}_1 \\
\mathbf{b}_2 \\
\mathbf{b}_3 \\
\end{bmatrix}
+
\begin{bmatrix}
\mathbf{Z}_1 & 0 & 0 \\
0 & \mathbf{Z}_2 & 0  \\
0 & 0 & \mathbf{Z}_3 \\
\end{bmatrix}
\begin{bmatrix}
\mathbf{u}_1 \\
\mathbf{u}_2 \\
\mathbf{u}_3 \\
\end{bmatrix}
+
\begin{bmatrix}
\boldsymbol{\epsilon}_1 \\
\boldsymbol{\epsilon}_2 \\
\boldsymbol{\epsilon}_3 \\
\end{bmatrix}
\end{align}

Moreover, as mentioned above, we assume unique variances and covariances for each trait/time point. 

\begin{align}
\textrm{var}
\begin{bmatrix}
\mathbf{u}_1 \\
\mathbf{u}_2 \\
\mathbf{u}_3 \\
\boldsymbol{\epsilon}_1 \\
\boldsymbol{\epsilon}_2 \\
\boldsymbol{\epsilon}_3 \\
\end{bmatrix}
&=
\begin{bmatrix}
\mathbf{G} \sigma^2_g{_{11}}  & \mathbf{G}\sigma^2_g{_{12}}  & \mathbf{G}\sigma^2_g{_{13}}  & 0 & 0 & 0\\
\mathbf{G} \sigma^2_g{_{21}} & \mathbf{G}\sigma^2_g{_{22}}  & \mathbf{G}\sigma^2_g{_{23}}  & 0 & 0 & 0\\
\mathbf{G} \sigma^2_g{_{31}}  & \mathbf{G}\sigma^2_g{_{32}}  & \mathbf{G}  \sigma^2_g{_{33}} & 0 & 0 & 0 \\
0 & 0 & 0 & \mathbf{I} \sigma^2_{\epsilon}{_{11}}  & \mathbf{I} \sigma^2_{\epsilon}{_{12}} &  \mathbf{I}  \sigma^2_{\epsilon}{_{13}} \\
0 & 0 & 0 & \mathbf{I} \sigma^2_{\epsilon}{_{21}} & \mathbf{I} \sigma^2_{\epsilon}{_{22}} & \mathbf{I} \sigma^2_{\epsilon}{_{23}} \\
0 & 0 & 0 & \mathbf{I} \sigma^2_{\epsilon}{_{31}} & \mathbf{I} \sigma^2_{\epsilon}{_{32}} & \mathbf{I} \sigma^2_{\epsilon}{_{33}} \\
\end{bmatrix}
\end{align}

Thus, for this simple case, we are capturing the full covariance across the three time points and leveraging this covariance to predict unique genetic values at each. However, notice the dimensions of the covariances $\sigma^2_g$ and $\sigma^2_{\epsilon}$. Here, we must solve for 12 parameters. If we have a very large population, this may not be an issue. However, if we consider a more realistic data set from HTP, it is likely that we will have many more time points. Thus, for $t$ time points, we will need to estimate $t$ variances and $t(t-1)/2$ covariances for both the genetic effects and residuals. For most HTP studies, this will create unnecessary computational demands. Moreover, additional challenges could be experienced if the parameter estimates are near the bounds, which may yield inaccurate estimates of variance components. Thus, when faced with larger longitudinal data sets ($t >$ 5), the researcher should question whether it is necessary to estimate each covariance. In cases in which the measurements are taken at short intervals within a given developmental period, it is likely safe to assume that the genetic variances between adjacent time points will be very similar. Therefore, a much simpler model may still capture much of the covariance while estimating fewer parameters. This is discussed in detail below. For other cases in which fewer measurements are recorded over more widely spaced intervals, the previous assumption may not hold true, and the full, unstructured matrix used in the multi-trait framework may be a more accurate model.

\subsection{Covariance functions and random regression models for longitudinal genetic prediction}
In the multi-trait framework, we treat the longitudinal phenotype, say growth, as a collection of independent traits; as a result, we are limited to making predictions at time points with records. However, in most longitudinal studies, we are interested in learning about the development of a continuous trait over time and do so by taking measurements at discrete time points. The time points or intervals themselves may be selected somewhat arbitrarily, and we seek to fill in information between time points. Thus, to capture the full trajectory of trait development, we can separate the trajectory into infinitely smaller intervals. Therefore, if we view the longitudinal trait as an ``infinite-dimensional" trait, we can see that the multivariate framework is inadequate, in that it does not directly consider the time axis and it does not allow us to make predictions at time points without observations. 

\citet{kirkpatrick1990analysis} initially proposed the use of covariance functions (CFs) for the analysis of ``infinite-dimensional" traits. A CF is simply the infinite-dimensional equivalent of a covariance matrix for a given number of time points. Using this approach, the covariance between any two records measured at given time points can be obtained using only the time points and some coefficients. For an ``infinite-dimensional" trait, there can be an infinite number of coefficients; however, in practice, the number of coefficients is dependent on the number of time points with records, with the maximum number of coefficients being $t(t + 1)/2$.

Following the example described in the multi-trait section above, we provide a brief example of the CF approach. Assume we have a trait measured at three time points using the covariance matrix in \cite{kirkpatrick1990analysis}. Using the multi-trait approach, we estimate the $3 \times 3$ additive genetic covariance matrix ($\hat{\boldsymbol{\Sigma}}$) and estimate the variances and covariances at each of the three time points. The goal of the approach described by \cite{kirkpatrick1990analysis} is to represent the additive generic covariance matrix ($\hat{\boldsymbol{\Sigma}}$) as a continuous covariance function ($\mathcal{K}$) given data collected at discrete time points. Although a number of methods can be used to estimate $\mathcal{K}$ from $\hat{\boldsymbol{\Sigma}}$, orthogonal polynomials are used most often due to the low correlations among the estimated coefficients \citep{schaeffer2004application}.

Given a covariance function with a full rank fit (e.g., order of polynomials is equal to the number of time points, $k = t$), \cite{kirkpatrick1990analysis} showed that the observed covariance matrix $\hat{\boldsymbol{\Sigma}}$ can be expressed as $\hat{\boldsymbol{\Sigma}} = \boldsymbol{\Phi} \mathbf{K} \boldsymbol{\Phi}'$, where $\mathbf{K}$ is a coefficient matrix associated with the CF, and $\boldsymbol{\Phi}$ is a matrix of Legendre polynomials of order $t$ by $k$, the order of Legendre polynomials (in this case $k = t$). $\boldsymbol{\Phi}$ is defined by the Legendre polynomial functions via $\boldsymbol{\Phi} = \boldsymbol{M}\boldsymbol{\Lambda}$. With Legendre polynomials, the time points are standardized so that they span an interval of -1 to 1, and here, $\boldsymbol{M}$ is a matrix of the polynomials of standardized time points. $\boldsymbol{\Lambda}$ is a matrix of coefficients of Legendre polynomials of order $k \times k$. The first two Legendre polynomials are $P_0(t) = 1$ and $P_1(t) = t$, and the subsequent $j^{th}$ Legendre polynomials are given by $P_{j+1}(t) = \frac{1}{j+1}(2j + 1) t P_j(t) - jP_{j-1}(t)$. These can be normalized to $\phi_j$ via $\phi_j = \frac{\sqrt{(2j + 1)}}{2} P_j(t)$. Thus, the first three normalized Legendre polynomials will be $P_0(t) = 0.707$, $P_1(t) = 1.2247t$, and $P_2(t) = -0.7906 + 2.3717t^2$. Thus, $\boldsymbol{\Lambda}$ is

\begin{align}
    \boldsymbol{\Lambda}
    &=
    \begin{bmatrix}
    0.7071 & 0 & -0.7906 \\
    0 & 1.2247 & 0 \\
    0 & 0 & 2.3717
    \end{bmatrix}
\end{align}

\noindent It is of particular importance to note that $\boldsymbol{\Phi}$ is not dependent on the values nor the time points in the data set; only $\mathbf{K}$ is. Thus, given the $3 \times 3$ covariance matrix, $\hat{\boldsymbol{\Sigma}}$, the covariance between any two time points, can be obtained using $\mathcal{K}(a_1, a_2) = \sum_{i = 0}^{\infty} \sum_{j = 0}^{\infty} \mathbf{K}_{ij} \phi_i(a_1) \phi_j(a_2)$, and the breeding value at any time point can be obtained using $\hat{g}_t = \sum_{i=0}^{k-1} \phi_i(d_t)u_i$. Moreover, with a full rank fit, the covariance matrix obtained is equivalent to that obtained using the multivariate approach in the previous section.

In most cases, the full covariance matrix $\hat{\boldsymbol{\Sigma}}$ is unknown; therefore, it must be estimated directly from the data. As shown by \cite{meyer1997estimation}, this can be done by a reparameterization of the multivariate or ``finite-dimensional" approach. However, in many studies, particularly those focused on the analysis of longitudinal milk production in dairy cattle, RR models (e.g., test-day models) are most commonly used. The RR approach proposed by \cite{schaeffer1994random} regresses the phenotypic trait trajectories directly on $\boldsymbol{\Phi}$ to estimate $\mathbf{K}$. As demonstrated by \cite{meyer1997estimation}, both the CF and RR approaches are equivalent. The general form of the RR model is 
\begin{align}
y_{tij} &= FE_i + \sum_{k=0}^{nf} \phi_{jtk} \beta_k + \sum_{k=0}^{nr} \phi_{jtk} u_{jk} + \sum_{k=0}^{nr} \phi_{jtk} pe_{jk} + \epsilon_{tij}
\end{align}
\noindent Here, $FE_i$ is the fixed effect for the $i^{th}$ group; $\phi_{jtk}$ is the $k^{th}$ Legendre polynomial for individual $j$ at time $t$; $\beta_k$ is the fixed regression coefficient for the $k^{th}$ Legendre polynomial, which represents the overall mean trait trajectory for the population or group; $u_{jk}$ is the genetic value for the $k^{th}$ Legendre polynomial for the $j^{th}$ individual; and $pe_{jk}$ is the permanent environmental effect for the $k^{th}$ Legendre polynomial for the $j^{th}$ individual. This permanent environmental effect is a stable, perpetual, non-genetic effect that influences an individual's trait trajectory. It is assumed to be common to all repeated observations on the same individual. Thus, $\mathbf{e}$ can be considered temporary environmental effects. In matrix form, the RR model can be written as $\mathbf{y} = \mathbf{Xb} + \mathbf{Za} + \mathbf{Qp} + \boldsymbol{\epsilon}$. 

In the examples above, we used a full-order polynomial to model the covariance across time points. As in the multivariate example, this requires estimation of a large number of parameters and in most cases is computationally unfeasible and could lead to convergence problems or inaccurate parameter estimates. In most cases, it is much more advantageous to fit a simpler model using a reduced-order polynomial ($k < t$). This effectively allows fewer parameters to be estimated while still adequately describing $\hat{\boldsymbol{\Sigma}}$. Generally speaking, the goodness of fit will increase as the number of function parameters describing the curve increases \citep{pool2000reduction}. \citet{campbell2018utilizing} used random regression models using Legendre polynomials for GS of rice shoot growth trajectories and demonstrated that the model could improve the accuracy of prediction compared to that of a single time point model.

\section{Conclusion}
This chapter described statistical methods for analyzing large-scale HTP data in quantitative genetics. 
We contend that the integration of HTP data into quantitative genetics models triggers a great leap forward in plant breeding. 
In particular, we discussed 1) the genetic gain that can be achieved using HTP data, 2) the use of HTP data as predictive covariates in GS models, and 3) the modeling of temporal HTP data using RR models. 
In GS, it is known that the accuracy of genomic prediction, and thus the response to selection, decreases as the selection cycle advances \citep{habier2007impact,jannink2010dynamics}. To maintain the response to selection, it is necessary to update the model on a regular basis \citep{jannink2010dynamics,iwata2011prospects,yabe2013potential}. In order to update the model, it is necessary to conduct a field test to measure phenotypes and to obtain genome-wide markers for many genotypes. At this step, phenotypic measurement for model updating may become a serious bottleneck of GS breeding. Thus, it is important to utilize HTP, which can evaluate many genotypes and possibly shorten the time required for selection. 

High-throughput phenotyping and phenomics offer numerous opportunities to understand plant development, the genetics of quantitative traits such as yield, and their connection to the environment. The utilization of HTP data that are correlated with traits of interest can change how breeders select their material for advancement. Incorporating HTP data into prediction models has the potential to increase prediction accuracy, thus enabling plant breeders to select and discard more accurately.  Although the reviewed studies considered different models, they concluded that regardless of the model configuration, the inclusion of HTP data increased the prediction performance when it was combined with different data types (marker genotype, pedigree, and environment). Additional gains can be expected when considering interactions with environmental factors. 

The RR approach offers several advantages compared to the multivariate approach. As mentioned above, the RR approach allows environmental effects to be partitioned into permanent and temporary environmental effects. Moreover, the RR approach models the individual-specific deviations from the mean phenotypic trajectories of the population. This allows the shape, amplitude, and intercept of the phenotypic trajectories to be unique for each individual and assumes that the genetic and permanent environmental effects are not constant throughout trait development. Thus, the RR model should more accurately reflect the biological processes that give rise to the phenotype. Furthermore, RR models offer a robust framework for fitting reduced-fit covariance functions. This offers a computational advantage over the multivariate approach in that it allows the model to converge more quickly. Moreover, by only estimating the parameters that are necessary to describe the data, sampling errors can be minimized. Finally, the RR approach provides a robust framework that allows the researcher to study how genetic variability changes over time and enables selection of individuals to alter phenotypic trajectories over time.

\clearpage
\newpage 
\bibliographystyle{apalike} 
\bibliography{MorotaHTP}

\end{document}